\documentclass[12pt,preprint]{aastex}
\usepackage{natbib}

\shorttitle{Globular Cluster Formation in M82}
\shortauthors{Lipscy \& Plavchan}

\begin{document}
\title{Globular Cluster Formation in M82}
\author{S.~J. Lipscy and P. Plavchan}
\affil{Department of Physics and Astronomy, University of California,\\
       Box 951547, Knudsen Hall, Los Angeles, CA 90095-1562; \\
       lipscy@astro.ucla.edu; plavchan@astro.ucla.edu}

\begin{abstract}
We present high resolution mid-infrared (mid-IR; 11.7 and 17.65
$\micron$) maps of the central 400 pc region of the starburst galaxy
M82.  Seven star forming clusters are identified which together
provide $\sim$ 15\% of the total mid-IR luminosity of the galaxy.
Combining the mid-IR data with thermal radio measurements and near-
and mid-IR line emission, we find that these young stellar clusters
have inferred masses and sizes comparable to globular clusters.  At
least 20\% of the star formation in M82 is found to occur in
super-star clusters.
\end{abstract}


\section{Introduction}

M82 (NGC 3034) is often considered the archetypical starburst galaxy
since it has a derived star formation rate ($\sim$10 M$_{\odot}$
yr$^{-1}$; \citealt{om78}) that would deplete the observed molecular
gas in $<$ 10$^8$ yrs (i.e. short on a Hubble timescale;
\citealt{lord96}).  The starburst phenomena traces recent star
formation since it is the massive stars ($>$8 M$_{\odot}$), which have
short lives, that dominate the energetic output of the host galaxy.
The nuclear starburst of M82 dominates the infrared (IR) luminosity of
the galaxy - essentially all of the galaxy's L$_{IR} \sim$ 3.6
$\times$ 10$^{10}$ L$_{\odot}$\footnote{\citet{telesco80} published
L$_{IR}$ = L(1-300$\micron$) = 3.0 $\times$ 10$^{10}$ L$_{\odot}$
assuming a distance to M82 of 3.3 Mpc.  We have updated L$_{IR}$ for
the distance of 3.6 Mpc \citep{freedman94, sakai99} used in this
paper.} comes from the central kpc \citep{telesco80}.

Optical and near-IR imaging of M82 has revealed numerous
super-star clusters in its active star-forming nucleus.  Ground-based
optical imaging of the central region detected eight young knots
distributed throughout the region \citep{om78}.  Further study of
these knots broke them into smaller star clusters (half light
diameters $\sim$3-4 pc) and suggested cluster dynamical masses
10$^4$-10$^6$ M$_{\odot}$ \citep{smith01, degrijs01, mccrady03}.

M82's current starburst is thought to have been triggered by its
interaction with M81 $\sim$10$^8$ yrs ago
\citep{cottrell77,achtermannlacy95}.  It has been argued that the
conditions resulting from interactions and mergers of galaxies are
favorable for globular cluster formation (eg. \citealt{taniguchi99}
and ref. therein).  UV studies of global properties of starbursting
galaxies have shown that as much as 20\% of the light is produced in
the luminous knots \citep{meurer95, zepf99}, suggesting a high
efficiency of cluster formation in starbursts.  Understanding the star
formation occurring in M82's nuclear region can provide insight into
both the general process of star formation in starburst environments
and also the process and efficiency of forming super-star
clusters.

The nearly edge-on geometry of M82 combined with heavy optical
extinction (A$_V \sim$ 5-25 mag; \citealt{lester90, telesco91,
larkin94, satyapal95}) has made the galaxy's central 400 pc difficult
to study.  Even in the near-IR where A$_{2.2 \micron} \sim$ 1/10A$_V$
\citep{rieke85}, it is difficult to directly measure the deepest star
forming regions.  The mid-IR region of the spectrum is essential
for probing deep into the central regions of M82 and revealing the
details of the intense star formation occurring in the central
regions.  Previously, the highest resolution mid-IR map of the central
region of M82 was a 12.4 $\micron$ map by \citet{telescogezari92},
which had a resolution $\sim$1.1$\arcsec$.  Mid-IR maps of M82 with
lower resolution have also been published by \citet{rieke80},
\citet{telesco89}, \citet{deitz89}, \citet{telesco91} and
\citet{forsterschreiber03}.  In this paper, we improve upon previous
observations by presenting higher resolution mid-IR maps of M82.  We
discuss evidence suggesting the sources in the maps are young
counterparts to globular clusters and estimate the efficiency of
super-star cluster formation in M82.

\section{Observations}

On 2003 April 23, we imaged M82 at 3.5 (L-band), 11.7, and 17.65
$\micron$ with the Long Wavelength Spectrograph (LWS;
\citealt{jones93}), a facility instrument at the W.M. Keck
Observatory.  LWS uses a 128 $\times$ 128 pixel Boeing Si:As detector
and has a plate scale of 0$\farcs$08 pixel$^{-1}$, resulting in a
10$\farcs$2 $\times$ 10$\farcs$2 field of view.  We used the
``chop-nod'' mode of observing, with a chop throw of 15$\arcsec$
north.  The bad pixels in the images have been smoothed over, and a
mask has been applied to remove the portion of the chip not
illuminated by the source.  The seeing varied during the course of the
observations so the resolution in individual frames ranged from
0.4-1.0$\arcsec$ at 11.7 $\micron$ and from 0.5-0.7$\arcsec$ at 17.65
$\micron$.  At each of seven pointings across M82's nuclear region, we
imaged at all three wavelengths before moving to the next pointing.
Images from the seven pointings were mosaiced by centroiding on the
bright sources in each frame.  Assigning coordinates to the field was
accomplished by aligning 2MASS sources with the centroids of the two
bright sources in the L-band frames.  We estimate that our positions
are accurate to $\sim$0$\farcs$5.  The star $\mu$ UMa was used as the
primary standard for flux density and point-spread function
calibration, and $\alpha$ Her and $\eta$ Sgr were used to estimate a 
calibration error of $\sim$ 20\%.  

\section{Characterization of mid-IR sources}

Our mid-IR maps, presented as Figures 1 and 2, contain several bright,
resolved sources as well as diffuse emission connecting the brighter
sources.  The mid-IR sources, labeled A-G from west to east, are
denoted by black circles (with radii 0$\farcs$5, corresponding to the
positional error).  For reference, the 2 $\micron$ peak is marked in
the Figures as a yellow cross \citep{deitz86} and the dynamical center
measured by \citealt{lester90} is marked with a yellow circle.  The
sharp edges visible in the maps (eg. southeast of source B in the
11.7 $\micron$ map) are artifacts that resulted from the mosaicing of 
images with differing thermal backgrounds and do not affect the results 
of this paper. 

The earlier, lower resolution 12.4 $\micron$ map by
\citet{telescogezari92} contained two bright regions, one to either
side of the galaxy's center.  The overall structure in our higher
resolution maps is comparable, though we identify individual resolved
sources within the Telesco \& Gezari sources. To compare their flux
densities with those previously published, Telesco \& Gezari reported
a flux of 6.5 $\pm$ 0.7 Jy at 12.4 $\micron$ for the 4$\arcsec \times$
4$\arcsec$ region containing our sources C and D.  We measure a flux
for the same region of 5.3 $\pm$ 1.3 Jy at 11.7 $\micron$, in
agreement with the previous observation.

The flux densities and half intensity major and minor axes measured
from the maps for the mid-IR sources are listed in Table 1.  For
sources covered by more than one image, photometry was performed on
each image individually and the average of the measurements is
presented here.  The standard deviation of the measurements due to
background variations was $\leq$ 15\% and, combined quadratically with
the calibration error, results in an overall flux density error of
$\leq$ 25\% in our measurements.  

Using our 11.7 $\micron$ and 17.65 $\micron$ flux densities, we
estimate the color temperature of the mid-IR emitting dust to be in
the range 150-270 K, assuming the dust particles radiate as
blackbodies.  The mid-IR luminosities (L$_{MIR} \equiv$ L$_{12-18
\micron}$) of the sources, estimated by fitting the measured flux
densities to a blackbody with the color temperatures for each source,
are all between 0.2-6 $\times$ 10$^{8}$ L$_{\odot}$ and sum to 2.4
$\times$ 10$^{9}$ L$_{\odot}$.  L$_{MIR}$ for the entire galaxy
estimated using the same procedure with the uncorrected IRAS 12 and 25
$\micron$ flux densities of 53 Jy and 274 Jy, respectively, gives 1.7
$\times$ 10$^{10}$ L$_{\odot}$, thus the seven mid-IR sources
contribute $\sim$15\% of M82's total L$_{MIR}$.

Using radio images at five frequencies, \citet{allen99} created
spectrally decomposed images of thermal (free-free) and non-thermal
(synchrotron) emission.  We find that not only do all the centers of
the thermal H II regions (marked in Figures 1 \& 2 with red squares)
match the centers of the mid-IR sources reasonably well, the diffuse
structure in the mid-IR maps follows closely the structure in the
thermal free-free map.  \citet{golla96} also noted the correspondence
of the diffuse 1.5 \& 22 GHz emission with the \citet{telescogezari92}
12.4 $\micron$ emission.  This supports the hypothesis that the mid-IR
sources are heavily obscured H II regions.

The [Ne II] (12.8 $\micron$) map published by
\citet{achtermannlacy95}, with a resolution of $\sim$1$\arcsec$, also
correlates well with our mid-IR maps (see Figure 3).  It should be
noted that while the peak of the [Ne II] line lies within our 11.7
$\micron$ filter bandpass ($\Delta\lambda$ = 2.4 \micron), based on
the 5-16.5 $\micron$ spectra presented in \citet{forsterschreiber03},
the [Ne II] line contributes $\leq$15\% of the flux in our bandpass.
We interpret the correlation between the mid-IR and the [Ne II]
emission as confirmation that the source of the mid-IR emission is
dust heated by UV from young stars which ionize Ne I.  Further
evidence to support this hypothesis comes from the Br$\gamma$ map from
\citet{larkin94}.  Though not covering the entire mid-IR field, the
Br$\gamma$ emission observed is also well correlated with the mid-IR
emission.  Both emission line maps agree with the mid-IR map in the
apparent lack of emission toward the dynamical center of M82.

Non-thermal radio sources from \citet{mcdonald02} and \citet{allen99}
are shown in Figures 1 \& 2 as magenta crosses; these are assumed
to be supernova remnants since they have inverted radio spectra.  We
find no correlation between the radio supernova remnants and the 
mid-IR sources, but notice that most of the supernova remnants follow
the outer edge of the mid-IR emission at a flux levels $\lesssim$0.3
Jy arcsec$^{-2}$ at 11.7 $\micron$ and $\lesssim$0.7 Jy arcsec$^{-2}$
at 17.65 $\micron$.

Following \citet{beck01}, we assume a reference OB star luminosity of
2.5 $\times$ 10$^5$ L$_{\odot}$ for an O7 star \citep{vacca96} to
calculate the OB star content of each mid-IR source from its
L$_{MIR}$.  These values are listed in Table 1.  For the brightest
regions (B, C, and G) we find that $\sim$2000-2500 O7 stars are
required to generate the mid-IR luminosity.
\citet{fosterschreiber01} derived comparable numbers of OB stars for
regions near sources B and C (3200 and 4400 O7 stars, respectively)
using He I/Br$\gamma$ line ratios to estimate the Lyman continuum flux
from their regions.

Since we find that the [Ne II] emission correlates spatially with the
mid-IR emission, we use [Ne II] channel maps from
\citet{achtermannlacy95}, which have a resolution of 16 km s$^{-1}$,
to estimate velocity dispersions ($\sigma_{vel}$) in the ionized gas
of $\sim$15-30 km s$^{-1}$ at the positions of the mid-IR sources.
Estimates of the velocity dispersions from the Br$\gamma$
\citep{larkin94} and $^{13}$CO \citep{neininger98} channel maps, each
with lower velocity resolution, give similar results.  A caveat to
using ionized emission to estimate a velocity dispersion is that the
ionized gas may well be accelerated by shocks, and therefore its
velocity should be considered an upper limit to the stellar velocity.

From the range of estimated velocity dispersions and assuming
virialized systems, we use $M = \eta \sigma_{vel}^2 r_{h}/G$, to
calculate a range for the total mass in each these systems.  In this
formulation, G is the gravitational constant, $r_{h}$ is the projected
half intensity radius defined to be the geometric mean of the
semi-major and semi-minor axes, and $\eta$ = 10 \citep{smith01}.  The
mass ranges for each mid-IR source are listed in Table 1 and should be
considered upper limits since H II regions are generally found to be
freely expanding systems.  For the largest sources, the mass range we
find is 6-25 $\times$ 10$^6$ M$_{\odot}$.  In every case, the range
contains the system mass found by extrapolating from the number of O7
stars using a Salpeter mass function.  Comparing the sizes and masses
of the star clusters forming in M82 to three of the largest Galactic
globular clusters, which have radii of 5-10 pc and masses 1-2 $\times$
10$^6$ M$_{\odot}$ (see Table 1), we conclude that the star clusters
forming in the nuclear region of M82 are young analogs to globular
clusters.

The existence of present day globular cluster-sized knots of star
formation is not unique to M82.  Indeed, super-star clusters
containing quantities of OB stars similar to those we find in M82 have
been observed in a number of galaxies (e.g. NCG 5253,
\citealt{turner03}; He 2-10, \citealt{johnson03}; NGC 4038/9,
\citealt{mengel02}).  Additionally, two star forming regions in the
Milky Way, the Arches cluster near the Galactic center and the Cygnus
OB2 association, are estimated to each weigh in at 6 $\times$ 10$^4$
M$_{\odot}$ \citep{serabyn98, knodlseder00}, comparable to a small
globular cluster \citep{mandushev91,pryor93}.  However, the Arches
cluster and Cygnus OB2 each contain around 100 O stars, which
contribute together only $\sim$3\% of the total Galactic O stars
\citep{terzian74}, and yet far outnumber all other Galactic star
forming regions in their O star content.

In M82, the seven mid-IR sources in the nuclear region together
contribute $\sim$ 15\% of the total L$_{MIR}$ of the galaxy.  The [Ne
II] map suggests that there may be several H II regions outside our
mid-IR field which may contribute up to an additional $\sim$5\% to
the total L$_{IR}$.  Assuming all the mid-IR luminosity in M82 comes
from star formation \citep{telesco88}, it follows that $\gtrsim$20\%
of M82's star formation is in the form of super-star clusters - in
contrast to the mere 3\% in the Milky Way.  This may be an important
feature of starbursts in general; not only do they provide an
environment suitable for forming globular clusters, but the super-star
cluster formation efficiency in starbursts is $\gtrsim$20\%.

\section{Conclusions}

This paper presents mid-IR (11.7 and 17.65 $\micron$) maps with
$\sim$0.5$\arcsec$ resolution of the central 400 pc of the
starbursting galaxy M82.  We find 7 resolved sources in this region of
M82 with luminosities summing to 15\% of the total IRAS flux of the
entire galaxy.  The mid-IR maps exhibit features comparable to those
found in maps of [Ne II] emission, Br$\gamma$ emission, and thermal
free-free emisson.  We present evidence implying that the mid-IR
sources are giant H II regions in which globular cluster-sized star
clusters are forming.  Our data imply that $\gtrsim$20\% of the star
formation in M82 is occurring in super-star clusters.

\acknowledgments This work has been supported by funding from NASA.
The authors wish to thank Mike Jura, Jean Turner, and James Larkin for
helpful comments.  Data presented herein were obtained at the
W.M. Keck Observatory, which is operated as a scientific partnership
among the California Institute of Technology, the University of
California and NASA. The Observatory was made possible by the generous
financial support of the W.M. Keck Foundation.  The authors wish to
recognize and acknowledge the very significant cultural role that the
summit of Mauna Kea has always had within the indigenous Hawaiian
community.  We are most fortunate to have the opportunity to conduct
observations from this mountain.  This publication makes use of data
products from the Two Micron All Sky Survey, which is a joint project
of the University of Massachusetts and the Infrared Processing and
Analysis Center/California Institute of Technology, funded by the NASA
and the National Science Foundation.

\clearpage

\begin{deluxetable}{ccccccccc}
\tablenum{1} 
\tablecaption{Properties of Proto-Globular Clusters in
M82 and Galactic Globular Clusters}
\label{table1}
\tablewidth{0pt}
\tablehead{
\colhead{Source} & \colhead{F$_{\nu}$(11.7 $\micron$)} & 
\colhead{F$_{\nu}$(17.65 $\micron$)} & \colhead{Size\tablenotemark{a}} & 
\colhead{T$_c$} & \colhead{L$_{MIR}$} & 
\colhead{N(O7)} & \colhead{Mass\tablenotemark{b}}\\
\colhead{} & \colhead{(Jy)} & 
\colhead{(Jy)} & \colhead{(pc)} & \colhead{(K)} & 
\colhead{(10$^8$ L$_{\odot}$)} & 
\colhead{} &\colhead{(10$^6$ M$_{\odot}$)}
}
\startdata
A & 0.16 & 0.15 & 10 $\times$ 10 & 270 & 0.20 &  80  & 0.3\tablenotemark{c} \\
B & 3.17 & 5.39 & 30 $\times$ 21 & 195 & 5.9  & 2400 & 6-25 \\
C & 2.38 & 5.95 & 26 $\times$ 19 & 165 & 5.5  & 2200 & 6-22 \\
D & 1.23 & 3.31 & 21 $\times$ 12 & 160 & 3.3  & 1300 & 4-16 \\
E & 0.67 & 1.74 & 13 $\times$ 13 & 160 & 1.7  & 700  & 3-13 \\
F & 0.77 & 2.24 & 28 $\times$ 8  & 155 & 2.3  &  900 & 4-15 \\
G & 1.56 & 4.96 & 25 $\times$ 16 & 150 & 4.9  & 2000 & 5-20 \\
$\omega$ Cen & & & 9\tablenotemark{d} & & & & 2\tablenotemark{e} \\
NGC 6273 & & & 3\tablenotemark{d} & & & & 1\tablenotemark{e} \\
NGC 104  & & & 5\tablenotemark{d} & & & & 1\tablenotemark{e} \\

\enddata
\tablenotetext{a}{Sizes are full width at half intensity major and minor 
axes of elliptical sources, as observed, and include a $\sim$0.5$\arcsec$ 
(9 pc) PSF.}
\tablenotetext{b}{Upper limit to mass of systems}
\tablenotetext{c}{Source A is not identifiable in the [Ne II] channel
maps.  Stellar mass extrapolated from the number of O7 stars using a
Salpeter initial mass function}
\tablenotetext{d}{Sizes are averages of values in \citet{vandenbergh91} and 
\citet{mandushev91}}
\tablenotetext{e}{From \citet{mandushev91}}

\end{deluxetable}

\clearpage

\figcaption[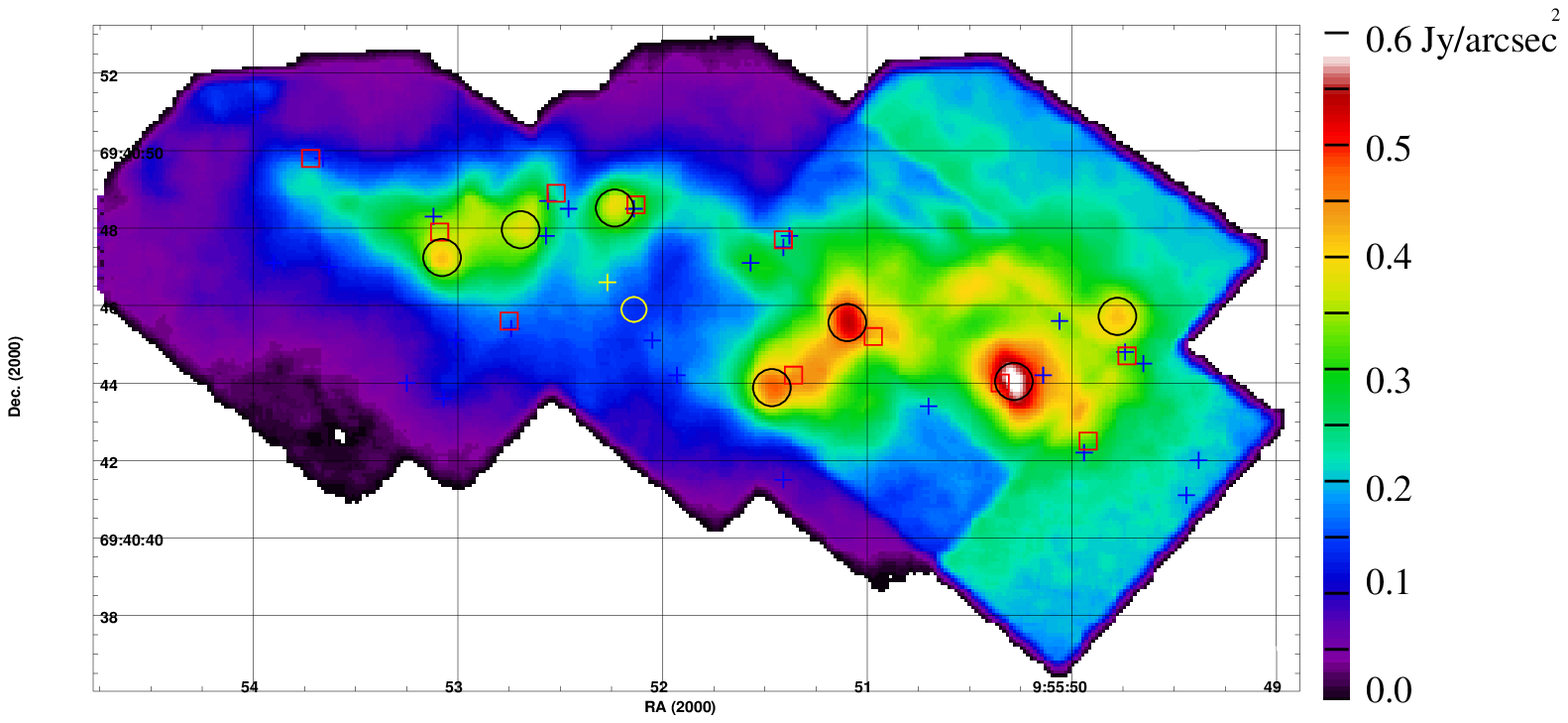]{Central region of M82 at
  11.7 $\micron$ smoothed with a 0$\farcs$4 boxcar function.  The
  seven mid-IR sources are labeled with white letters and their
  positional errors are represented by 0$\farcs$5 radius black circles.
  The 2 $\micron$ peak \citep{deitz86} is labeled with a yellow cross
  and M82's dynamical center as determined by \citet{lester90} is
  marked by a yellow circle. Magenta crosses mark positions for
  non-thermal radio sources \citep{mcdonald02, allen99} and red
  squares mark position of H II regions \citep{allen99}.}

\figcaption[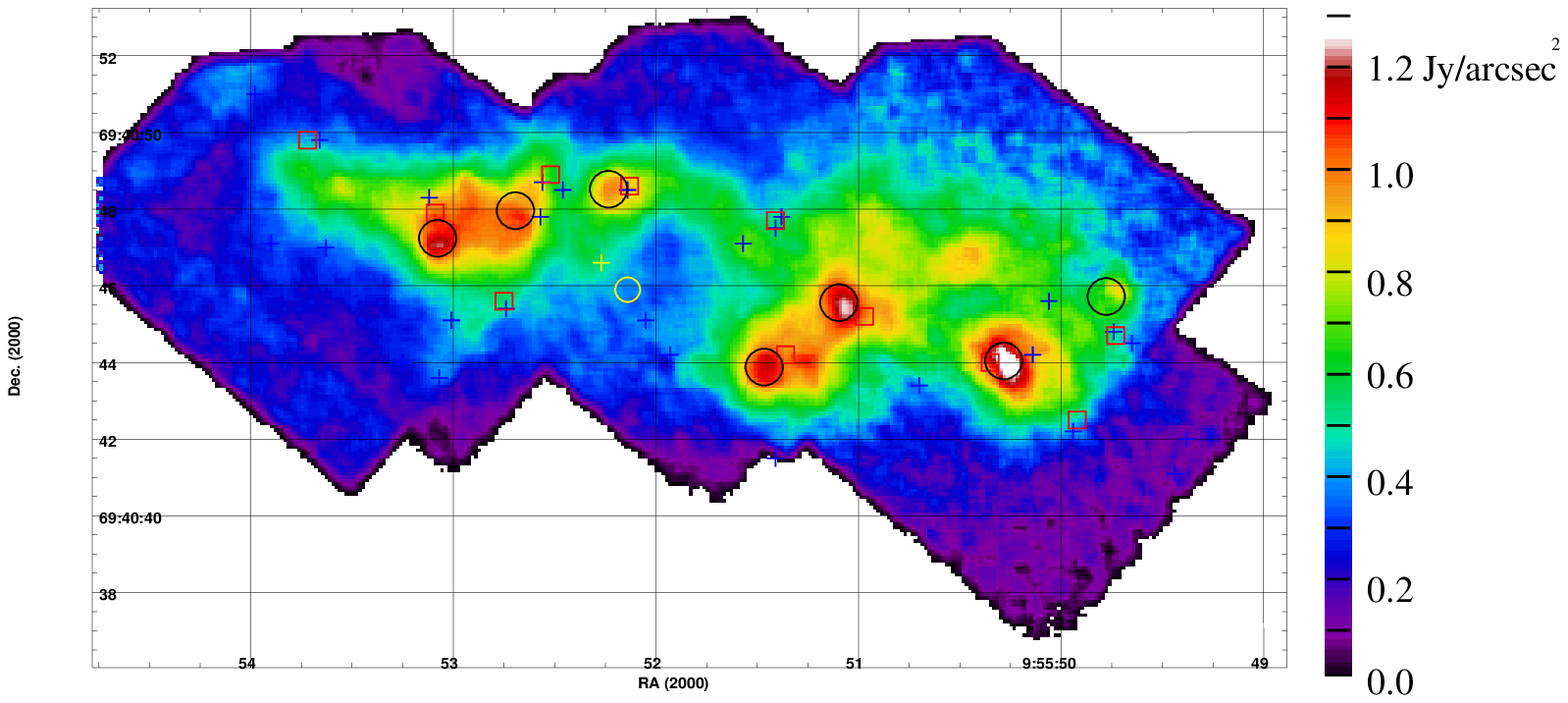]{Central region of M82 at
  17.65 $\micron$ smoothed with a 0$\farcs$4 boxcar function.  Symbols
  are the same as in Figure 1.}

\figcaption[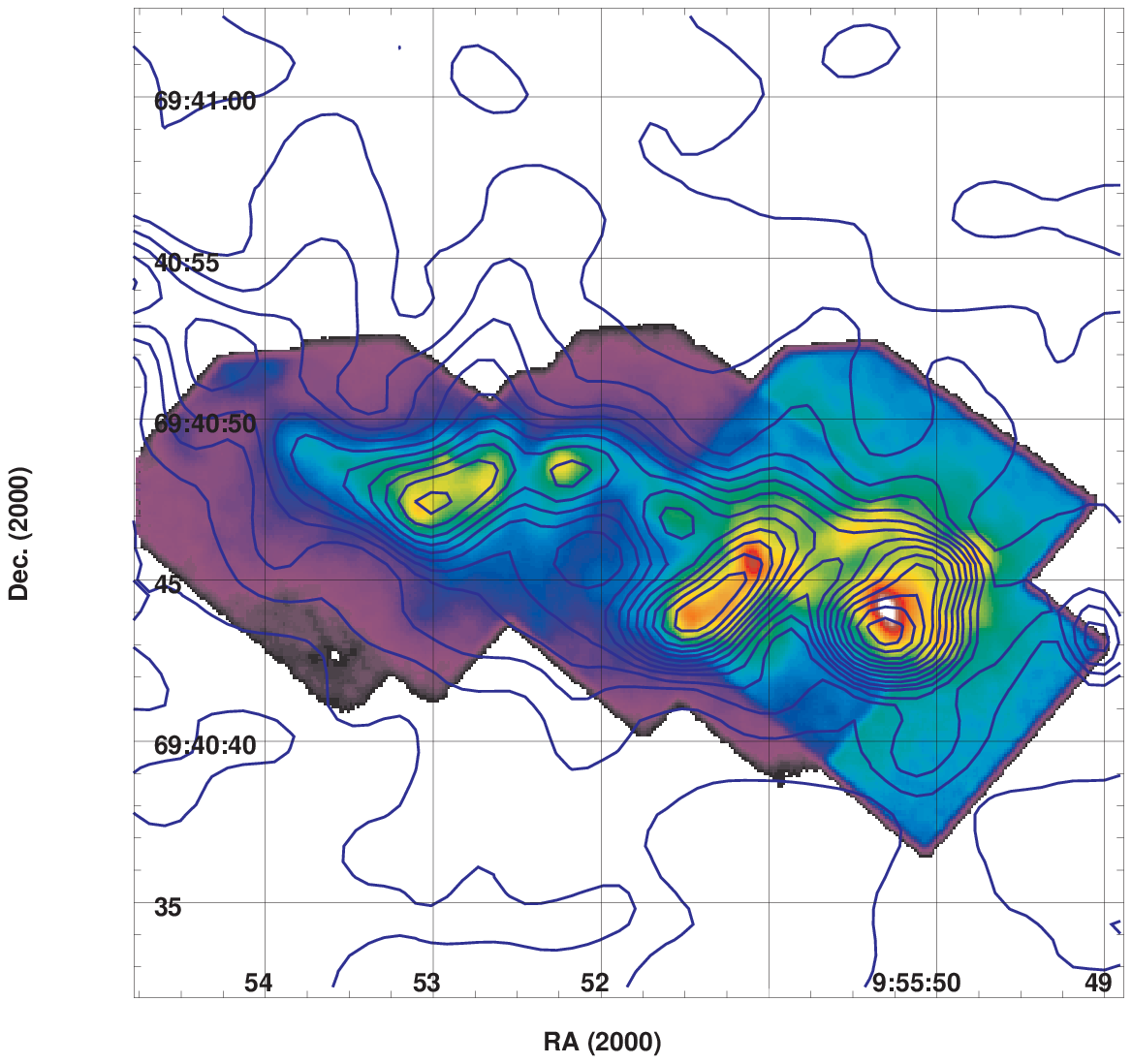]{Smoothed 11.7 $\micron$ data overlaid with
  contours from the \citet{achtermannlacy95} [Ne II] line map deconvolved
  with the maximum entropy method (their Figure 5).}

\clearpage

\begin{figure}
\plotone{f1.eps}
\end{figure}

\begin{figure}
\plotone{f2.eps}
\end{figure}

\begin{figure}
\plotone{f3.eps}
\end{figure}

\end{document}